\documentclass{svproc}
\usepackage{graphicx}%
\usepackage{multirow}%
\usepackage{amsmath,amssymb,amsfonts}%
\usepackage{mathrsfs}%
\usepackage[title]{appendix}%
\usepackage{xcolor}%
\usepackage{textcomp}%
\usepackage{manyfoot}%
\usepackage{booktabs}%
\usepackage{algorithm}%
\usepackage{algorithmicx}%
\usepackage{algpseudocode}%
\usepackage{listings}%
\usepackage{hyperref}
\usepackage{caption}
\usepackage{tabularray}
\usepackage{url}

\usepackage{booktabs}
\usepackage{array}
\definecolor{entitycolor}{RGB}{0, 0, 255} 
\definecolor{organizationcolor}{RGB}{255, 0, 0} 
\definecolor{conditioncolor}{RGB}{128, 0, 128} 
\definecolor{personcolor}{RGB}{0, 128, 0} 
\definecolor{conceptcolor}{RGB}{165, 42, 42} 
\definecolor{othercolor}{RGB}{255, 165, 0} 
\usepackage[utf8]{inputenc}

\begin{document}
\mainmatter              
\title{
\vspace{-6mm}
A New Perspective on ADHD Research: Knowledge Graph Construction with LLMs and Network Based Insights
\vspace{-3mm}
}
\titlerunning{LLM-Driven ADHD KG \& Network Insights}  
\author{
\vspace{-3mm}
Hakan T. Otal\inst{1}, Stephen V. Faraone\inst{2} \and M. Abdullah Canbaz\inst{1}
}


\authorrunning{Otal et al.} 

\institute{
Department of Information Science and Technology \\
College of Emergency Preparedness, Homeland Security and Cybersecurity \\
University at Albany, SUNY, NY, USA\\
\email{hotal@albany.edu}, \email{mcanbaz@albany.edu}\\
\and
Department of Psychiatry \\
Norton College of Medicine \\
Upstate Medical University SUNY, NY, USA\\
\email{sfaraone@childpsychresearch.org}}

\maketitle              

\vspace{-3mm}

\begin{abstract}
Attention-Deficit/Hyperactivity Disorder (ADHD) is a challenging disorder to study due to its complex symptomatology and diverse contributing factors. To explore how we can gain deeper insights on this topic, we performed a network analysis on a comprehensive knowledge graph (KG) of ADHD, constructed by integrating scientific literature and clinical data with the help of cutting-edge large language models. The analysis, including k-core techniques, identified critical nodes and relationships that are central to understanding the disorder. Building on these findings, we curated a knowledge graph that is usable in a context-aware chatbot (Graph-RAG) with Large Language Models (LLMs), enabling accurate and informed interactions. Our knowledge graph not only advances the understanding of ADHD but also provides a powerful tool for research and clinical applications.

\keywords{ADHD, Knowledge Graph, Network Analysis, Large Language Models, Retrieval-Augmented Generation}
\end{abstract}

\vspace{-8mm}
\section{Introduction}
\vspace{-2mm}

ADHD is a neurodevelopmental disorder characterized by persistent patterns of inattention, hyperactivity, and impulsivity that are disruptive and inappropriate for an individual’s developmental level. ADHD is one of the most common psychiatric disorders in children, with symptoms often continuing into adulthood \cite{Faraone2024}. The disorder has a complex etiology, influenced by a combination of genetic, neurobiological, and environmental factors. Despite extensive research, the precise mechanisms underlying ADHD remain poorly understood, partly due to its heterogeneous nature \cite{Faraone2024,hayashi2018}.

The complexity of ADHD is reflected in its diverse symptomatology, comorbidities, and the wide range of cognitive, behavioral, and social outcomes observed in affected individuals. Traditional approaches in ADHD research have typically focused on identifying specific deficits or abnormalities within isolated domains, such as neuroimaging or genetic studies \cite{barke2023}. While these studies have provided valuable insights, they sometimes fail to capture the intricate interconnections between different biological, cognitive, and behavioral aspects of this medical condition.

In parallel, network analysis, rooted in graph theory and extensively utilized across disciplines such as sociology, biology, and computer science, offers a robust framework for examining complex systems. By representing these systems as networks composed of nodes (e.g., brain regions, genes, symptoms) and edges (relationships between them), network analysis facilitates the exploration of the intricate interactions and dependencies within the system as a whole. This approach is particularly advantageous for studying multifactorial conditions like ADHD, where the interplay between various biological, genetic, and environmental factors is crucial to understanding the disorder’s etiology and manifestation.

Knowledge graphs (KGs) have become essential in network analysis, especially with the rise of Large Language Models (LLMs) and Retrieval Augmented Generation (RAG) systems. A KG is a structured representation that organizes data into triplets—head entity, relationship, and tail entity—forming a graph where entities are nodes and relationships are edges. This allows for the representation of complex, multi-hop interactions. In ADHD research, KGs integrate diverse data sources such as genetics, neuroimaging, and clinical symptoms, providing a comprehensive view of the disorder. Our approach builds on existing ADHD-related KGs~\cite{li_multi_hop_2023,jin_improving_2023,emmanuelpapadakis} by incorporating LLM-based insights and clinical data. Additionally, KGs help address hallucination issues in LLMs, where generated outputs are coherent but lack factual grounding. By using KGs as a structured, evidence-based foundation, LLMs can align their outputs with factual data, reducing hallucinations and improving the reliability of AI-generated insights. This is particularly important in medical and psychological domains, where accuracy directly affects clinical decisions and patient outcomes\cite{guo_knowledgenavigator_2024}.

The primary objective of this paper is to address the need for a comprehensive and structured representation of ADHD knowledge by leveraging Large Language Models (LLMs) to construct a multimodal KG specifically focused on ADHD. This study seeks to integrate information from a variety of sources, including scientific literature, clinical data, and expert knowledge, to create a rich and interconnected representation of ADHD. By systematically organizing and analyzing this knowledge using network-based approaches, the paper aims to achieve several key goals:
\vspace{-2mm}
\begin{itemize}
\item \textbf{Build a Ground-Truth Evidence Base:} Collect and validate expert-curated data to establish a reliable foundation for the ADHD KG. This evidence base will serve as a robust resource for downstream analysis and support further research and clinical applications.
\item \textbf{Support Decision-Making and Clinical Practice:} Demonstrate how the ADHD KG can be utilized as an evidence-based Retrieval Augmented Generation (RAG) system for an ADHD-Expert LLM. This application aims to create a ‘proof of concept’ expert system that shows how curated evidence has the potential to educate users, enhance decision-making and patient care by providing patients, clinicians and researchers with accurate, well-structured, and comprehensive information about ADHD.
\end{itemize}

\vspace{-7mm}
\section{Methodology}

\vspace{-3mm}
\subsection{Data Collection and Preprocessing}

This paper's foundation rests substantially on the work of Stephen V. Faraone, Ph.D., a Distinguished Professor at SUNY Upstate Medical University and a preeminent figure in ADHD research. Dr. Faraone's extensive body of work, encompassing psychiatric genetics, childhood mental disorders, and psychopharmacology, has been instrumental in shaping our current understanding of ADHD. His recent application of advanced machine learning techniques to these domains has further expanded the frontiers of ADHD research\cite{faraone2019,faraone2021}.

Dr. Faraone spearheaded the identification of the first genes associated with ADHD, significantly influenced the diagnostic criteria for adult ADHD, and established the ADHD Molecular Genetics Network in the 1990s\cite{faraone2005}. His research output, comprising over 1000 publications with an exceptional citation count of 195,382, underscores the impact and relevance of his work\cite{faraone2021,sunyupstate}. His provision of scientific papers, resources, and empirical data related to ADHD has been crucial in establishing the evidence base and ground truth information for our study. 

To create a preliminary, proof of concept, model, Dr. Faraone provided several review articles and other files of information that he had curated for this project. The input dataset contains manually curated 224 pages of information about ADHD spread across 15 files. To effectively process these resources, we employed the Langchain library in Python to load the directory of our documents. These documents were then segmented into smaller text chunks, each consisting of 1,500 characters with an overlap of 150 characters. This preprocessing step ensured that the text segments were both manageable and contextually coherent, thereby facilitating more accurate and effective downstream analysis.

\vspace{-3mm}
\subsection{Semantic Concept Extraction with LLM}
\vspace{-2mm}

To extract relevant concepts from the text chunks, we employed the Llama3.1-8B \cite{dubey2024llama3herdmodels} model, a state-of-the-art Large Language Model (LLM). The model’s primary function was to identify and categorize key concepts within the text, subsequently organizing these concepts into a structured graph where nodes represent the concepts and edges denote the relationships between them.

The extracted concepts were categorized into predefined types. Concepts that did not align with these categories were classified under the label ‘other’. 

\vspace{-3mm}
\subsection{Graph Construction and Refinement}
\vspace{-2mm}

The graph construction process began with the creation of an initial graph using the NetworkX library. In this graph, nodes represented the extracted concepts, while edges denoted the relationships between these concepts. Each edge was enriched with attributes such as edge\_type, edge\_details, weight, and a reference to the originating text chunk, ensuring that relationships were contextually grounded and traceable back to their source.

Direct relationship edges represent explicit semantic links between two concepts, while contextual proximity edges capture co-occurrences within the same document. For instance, a direct relationship would link ‘ADHD’ to ‘impulsivity,’ whereas a contextual proximity edge might link ‘ADHD’ to ‘neuroimaging studies’ if they frequently co-occur in the literature.

\vspace{-4mm}
\subsubsection{Contextual Proximity Edges:}
To further enrich the graph, we calculated contextual proximity between concepts. Concepts that appeared within the same text chunk were connected by edges, with edge weights reflecting the frequency of co-occurrence. This approach allowed us to capture the strength of associations between concepts, highlighting their contextual relevance within the broader network.

\vspace{-4mm}
\subsubsection{Eliminating Redundant Nodes:}
In the refinement stage, we addressed redundancy within the graph. The redundancy comes from the very similar or same meaning nodes being represented as different nodes in the graph. For example the node 'ADHD' and 'Attention-deficit/hyperactivity disorder' can be represented by two different nodes even though they mean the same thing. 

By generating embeddings for each node using the “gte-base-en-v1.5” transformer model\cite{zhang2024mgte}, we managed to represent each node as a vector (The embedding model was selected from the embedding models leaderboard\cite{muennighoff2022mteb} based on its strong performance and medium size [$\leq$1B parameters]). Using the DBSCAN clustering algorithm\cite{dbscan} with a very high similarity threshold (0.9), we identified clusters of nodes that were semantically very similar. The reason we set the threshold very high is we only needed to cluster very similar meaning nodes. So, these clusters enabled the detection and merging of redundant nodes, such as synonymous terms and abbreviations. In the final graph, each cluster was represented by the most relevant (highest-degree) node, with edges updated accordingly to reflect this consolidation, thereby reducing redundancy and enhancing the clarity and utility of the graph.

\vspace{-3mm}
\section{Experimental Results}
\vspace{-2mm}

In this section, we present the results of our network analysis on the constructed ADHD KG. Unlike typical network graphs, this KG is characterized by links that carry semantic meanings, representing relationships between different concepts. As such, the network science metrics reported here must be understood in the context of a graph where edges do not merely signify connections but also encode specific, meaningful relationships.

\vspace{-3mm}
\subsection{ADHD Knowledge as Multi-Layer Network}

\begin{figure}[!hb]
\vspace{-7mm}
\centering
\includegraphics[width=\linewidth]{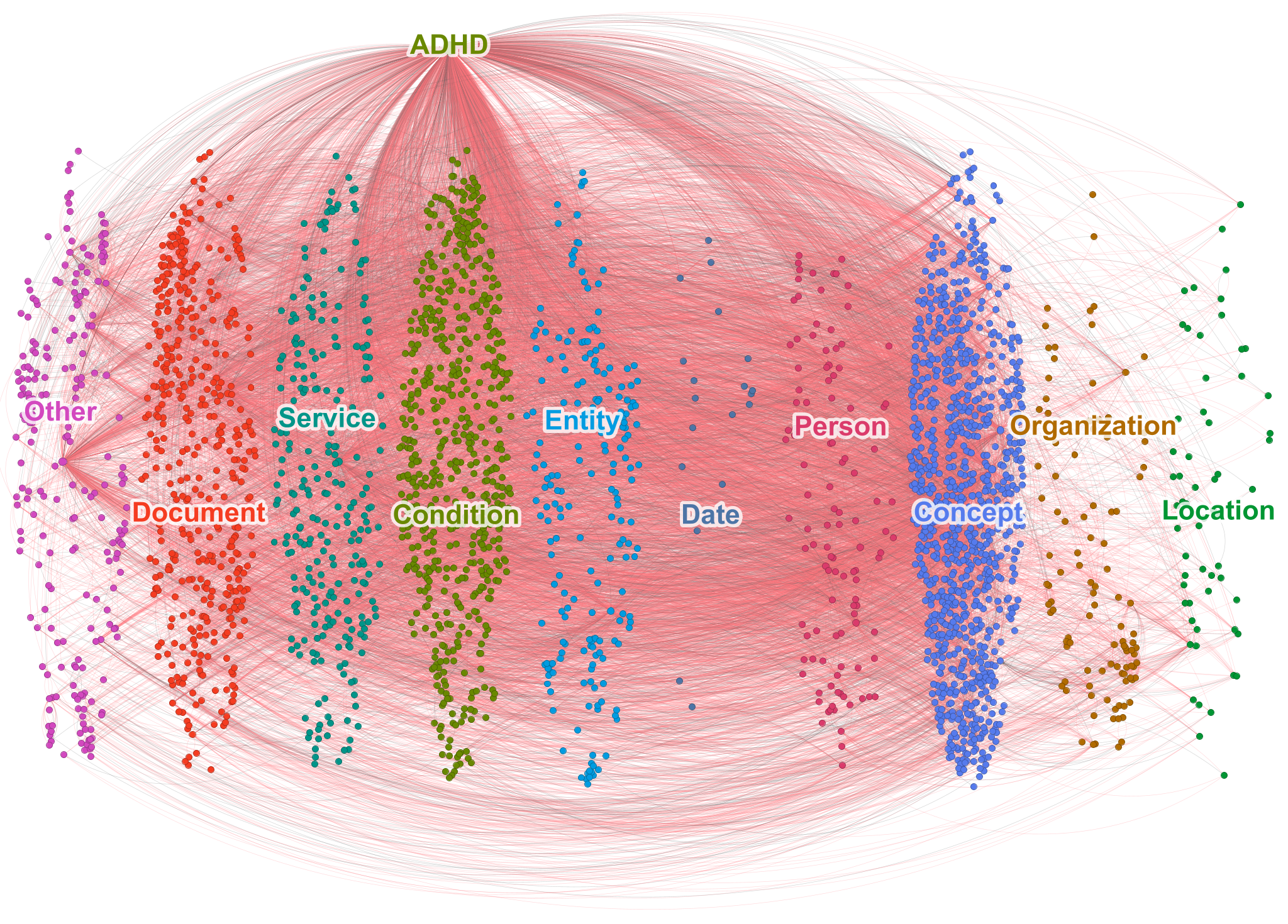}
\caption{ADHD Knowledge Graph Visualization (interactive version is at\\ \url{https://github.com/AI-in-Complex-Systems-Lab/ADHD-KnowledgeGraph})}
\label{fig:ADHD_network}
\vspace{-3mm}
\end{figure}

Figure~\ref{fig:ADHD_network} is a visualization of our KG centered around the concept of ADHD, with nodes representing various entities and concepts and edges illustrating the relationships between these nodes. The node labeled “ADHD” is prominently positioned at the top of the graph, serving as a central hub. This placement indicates that ADHD is the primary subject being explored, with numerous connections radiating out to other entities across different categories. The dense web of relationships around ADHD suggests a complex and multifaceted nature, with significant connections to a wide array of entities.

Colors are assigned to nodes based on their types, such as “Person,” “Organization,” “Concept,” “Documents,” and so forth. This color-coding helps to visually distinguish between different types of entities within the KG, making it easier to identify patterns and relationships within and across categories. The diverse coloring highlights how various aspects of ADHD are interconnected, providing a clearer understanding of how different entities relate to the central concept.

Edges between nodes are also color-coded based on their types. Red edges represent “relation” edges, indicating direct relationships between entities, such as a service provided by an organization or a document discussing a particular concept related to ADHD. In contrast, gray edges represent “contextual proximity” edges, suggesting that the entities are connected within a certain context, though not necessarily through a direct or explicit relationship. This might indicate shared attributes, common references, or similar thematic content, providing a richer context around the more explicit connections.

The categorization of nodes, such as “Document” “Service” and “Condition” shows the broad scope of ADHD’s influence. For instance, the connections between ADHD and “Condition” likely represent various medical or psychological conditions associated with ADHD, while “Service” might represent healthcare or educational services designed for individuals with ADHD. Similarly, “Document” likely refer to research papers, guidelines, or policy documents that discuss ADHD or its related aspects, creating a knowledge base that informs the other categories.

KG was created using Gephi software \cite{gephi}, a popular open-source tool for network analysis and visualization. Gephi allows for the assignment of colors to nodes based on their types and the coloring of edges according to their relationship type, as seen in this graph. This enhances the visual representation and makes it easier to interpret the complex relationships within the dataset. The hierarchical or layered structure of the graph, with “ADHD” at the top and various categories spreading out below, suggests a comprehensive approach to understanding how ADHD interacts with different aspects of society, healthcare, and research.

The less dense connections in categories like “Organization” and “Location” could suggest that while these entities are related to ADHD, they might play more specialized or niche roles compared to other categories. This visualization offers valuable insights into the broad impacts of ADHD and could be instrumental in identifying key areas of focus or gaps in research and understanding. For instance, if certain categories or nodes have fewer connections, it could indicate areas that need further exploration or development. The graph can serve as a powerful tool for researchers, clinicians, or policymakers to better understand the complex network of entities related to ADHD and to develop targeted interventions or policies accordingly.

\vspace{-3mm}
\subsection{Network Metrics}

\begin{table}[!hb]
\vspace{-8mm}
\centering
\small 
\begin{minipage}[!t]{0.48\linewidth} 
    \centering
    \caption{Network Metrics and \\Community Detection Results \\(filtered by node-degree $k_i>1$)}
    \begin{tblr}{
      column{2} = {r},
      hline{1,17} = {-}{0.08em},
      hline{2,11,13,15} = {-}{0.05em},
      rowsep = 2pt 
    }
\textbf{Metric}                         & \textbf{Value} \\
Nodes                                   & 2347           \\
Edges                                   & 8655           \\
Density                                 & 0.0031         \\
Avg. Degree                             & 6.289          \\
Triadic Closure                         & 0.0505         \\
Clust. Coef.                            & 0.8349         \\
Node Sim.                               & 0.0696         \\
Eig. Cent.                              & 0.0461         \\
Assort. Coef.                           & -0.1242        \\
Leiden Mod.                             & 0.6195         \\
Leiden Comm.                            & 31             \\
Louvain Mod.                            & 0.6157         \\
Louvain Comm.                           & 26             \\
G-N Mod.                                & 0.5876         \\
G-N Comm.                               & 74             
    \end{tblr}
    
    \label{tab:metrics}
\end{minipage}
\hfill
\begin{minipage}[!t]{0.48\linewidth} 
    \centering
    \caption{Number of nodes and edges grouped based on their types}
    \begin{tblr}{
      column{2} = {r},
      hline{1,18} = {-}{0.08em},
      hline{2,14} = {-}{0.05em},
      rowsep = 2pt 
    }
\textbf{Type}            & \textbf{Count} \\
\textbf{Node}            &                \\
concept                  & 904            \\
condition                & 579            \\
documents                & 434            \\
service                  & 268            \\
other                    & 221            \\
entity                   & 207            \\
person                   & 115            \\
organization             & 105            \\
location                 & 61             \\
date                     & 20             \\
\textbf{total \#nodes}    & 2914           \\
\textbf{Edge}            &                \\
contextual proximity     & 15289          \\
direct relationship      & 3640           \\
\textbf{total \#edges}    & 18929          
    \end{tblr}
    \label{tab:distributions}
\end{minipage}

\end{table}

Table \ref{tab:metrics} summarizes the key network metrics and community detection results obtained from the analysis. The KG comprises 2,347 nodes and 8,655 edges, reflecting the rich and varied relationships within the ADHD domain. With a network density of 0.0031, the graph is relatively sparse, indicating that while there are numerous relationships, they are selectively formed based on the specific meanings encoded in the edges.

The average degree of 6.289 suggests that each concept, on average, is connected to about six other concepts. This level of connectivity highlights the nuanced interplay between various entities, conditions, and other elements within the ADHD KG. The triadic closure value of 0.0505, together with an average clustering coefficient of 0.8349, points to a graph structure where concepts are likely to form tightly-knit groups based on their relationships, consistent with the idea that certain concepts in ADHD are more closely related or co-occur more frequently in the literature.

The relatively low average node similarity (0.0696) and eigenvector centrality (0.0461) reflect the diversity of concepts in the KG, where not all nodes are uniformly central or similar across the network. The assortativity coefficient of -0.1242 suggests a slight disassortative mixing pattern, where nodes of differing degrees tend to connect, potentially indicating that more specific or rare concepts may be linked to more general or well-established ones.


\vspace{-3mm}
\subsection{Community Detection}
\vspace{-2mm}

To further analyze the structure of the ADHD KG, we applied several widely used community detection algorithms. Community detection is essential in identifying groups of closely connected nodes, which can reveal the underlying structure of the graph and help in understanding how related concepts cluster together within the knowledge domain.

The Leiden algorithm \cite{Traag_2019} is a hierarchical clustering method that improves modularity by moving nodes between communities, yielding a score of 0.6195 and identifying 31 communities in our analysis. The Louvain algorithm \cite{Blondel_2008}, which also optimizes modularity, produced a similar score of 0.6157 with 26 communities, showing consistency in the community structure. In contrast, the Girvan-Newman algorithm \cite{Newman_2004} uses a divisive approach by removing high-betweenness edges, identifying 74 communities with a lower modularity score of 0.5876, indicating a finer partition of the graph.

The identified communities provide a structural view of how different research areas or clinical aspects of ADHD are connected. For example, some of the prominent clusters include:
\begin{itemize}
    \item Comorbidities: Conditions such as major depression, schizophrenia, autism, and obesity formed significant clusters, reflecting their frequent co-occurrence with ADHD in both clinical and genetic studies.
    \item Cognitive and Educational Outcomes: Concepts like intelligence, years of schooling, and subjective well-being also formed central communities, highlighting the impact of ADHD on cognitive development and life outcomes.
    \item Neuroimaging and Genetic Research: Research methodologies such as fMRI studies and concepts related to genetic correlation clustered together, demonstrating the focus on biological underpinnings and objective measurements of ADHD-related traits.
\end{itemize}
These communities reflect well-established areas of ADHD research, such as its genetic basis and neurodevelopmental effects, while also hinting at more specific areas of focus, such as the educational implications of the disorder.

Community detection revealed that key clusters related to ADHD comorbidities and educational outcomes. However, there is room for further exploration of community overlap with genetic factors. Moreover, this community information is valuable for various Graph-RAG (Retrieval Augmented Generation) applications. For instance, when a user queries the system with a specific question related to ADHD, the model can leverage community information to efficiently retrieve the most relevant subgraphs or concepts \cite{edge_local_2024}. This targeted retrieval can enhance the accuracy and context-specificity of the generated responses.

\vspace{-3mm}
\subsection{Node and Edge Type Distributions}
\vspace{-2mm}

Table \ref{tab:distributions} provides a detailed breakdown of the types of nodes and edges present within the KG. The node distribution reflects the diverse range of concepts captured, with ‘concept’ nodes being the most prevalent (904 instances), followed by ‘condition’ nodes (579 instances) and ‘documents’ nodes (434 instances). This variety indicates the graph’s comprehensive nature, encompassing various aspects of ADHD research, from clinical conditions to associated documents and services.

The edge type distribution is particularly telling of the KG’s distinct nature. ‘Contextual proximity’ edges are the most common, with 15,289 instances, indicating the frequency with which concepts co-occur within the same text chunks. This suggests that the graph is densely woven with contextual relationships. Additionally, there are 3,640 ‘direct relationship’ edges, which denote explicit connections between concepts as extracted by the LLM, highlighting the direct, semantically meaningful relationships present in the graph.

The low density of the KG suggests selective connections that highlight critical relationships rather than an overly connected or cluttered graph. The distribution of node and edge types emphasizes the unique structure of the ADHD KG, where each link carries a specific meaning, thus providing a rich resource for further analysis and understanding of ADHD’s complex web of related concepts.

\vspace{-3mm}
\subsection{K-Core Analysis}
\vspace{-2mm}

The k-core analysis of the ADHD KG, constructed using the evidence base provided by Dr. Faraone, offers several significant insights into the structure and focus areas of the graph. The nodes that persist in the maximum k-core (16 cores) reveal the most interconnected and central concepts, underscoring their importance in the study of ADHD.
\begin{figure}[!b]
\vspace{-4mm}
\centering
\begin{minipage}[!t]{0.57\linewidth} 
    \centering
    \includegraphics[width=\linewidth]{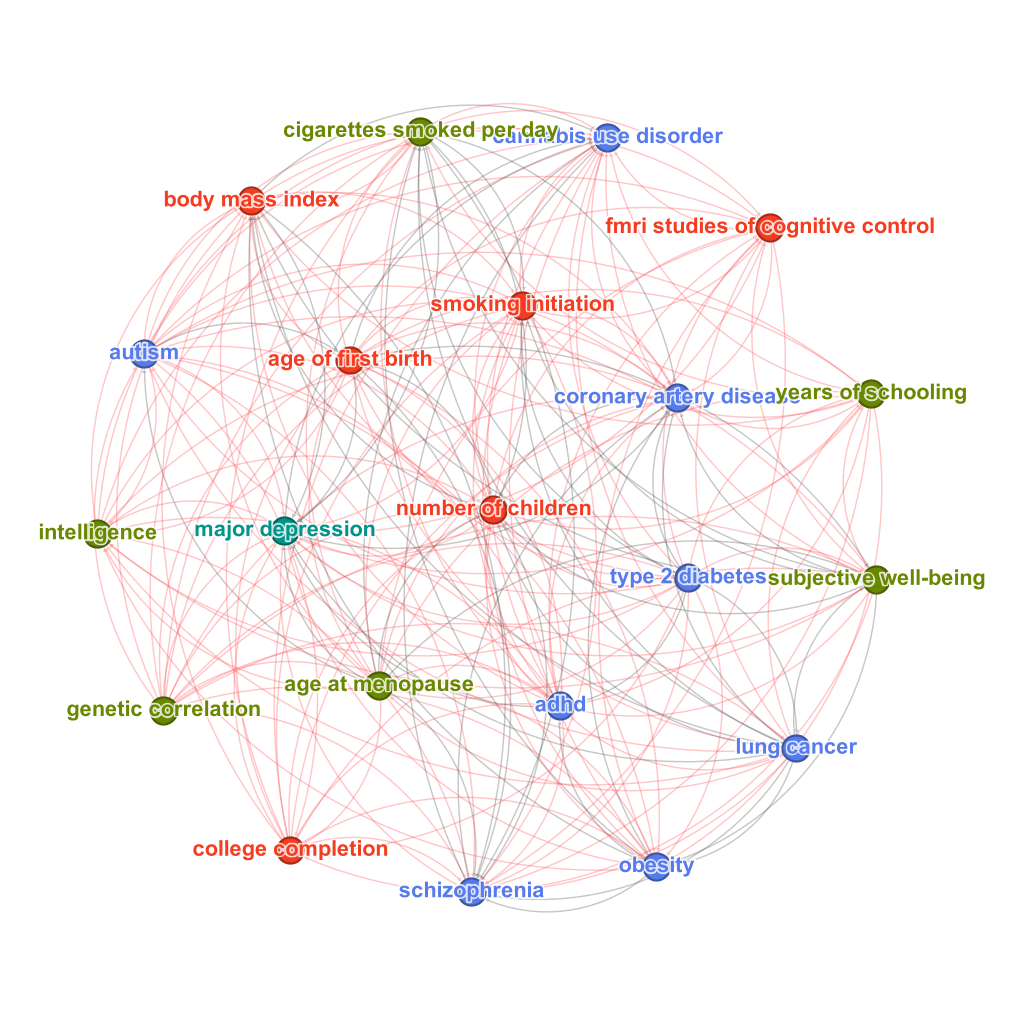} 
\end{minipage}
\hfill
\begin{minipage}[!t]{0.42\linewidth} 
    \vspace{0pt} 
    \centering
    \small
    \begin{tabular}{p{0.3\linewidth} p{0.7\linewidth}} 
    \toprule
    \textbf{Type} & \textbf{Nodes} \\ 
    \midrule
    \centering Concept & Years of schooling, \par Cigarettes smoked/day, \par Age at menopause, \par Subjective well-being, \par Genetic correlation, \par Intelligence \\
    \midrule
    \centering Condition & Lung cancer, \par Autism, \par Cannabis use disorder, \par Schizophrenia, \par Obesity, \par ADHD, \par Coronary artery disease, \par Type 2 diabetes \\
    \midrule
    \centering Documents & Number of children, \par FMRI studies, \par College completion, \par Body mass index, \par Age of first birth, \par Smoking initiation \\
    \midrule
    \centering Entity & Major depression \\
    \bottomrule
    \end{tabular}
\end{minipage}
\caption{k-core Network (K=16) on the left, and Nodes in the Maximum k-core (16 Cores) on the right.}
\label{fig:kcore_combined}
\vspace{-4mm}
\end{figure}

One key insight is the \textbf{centrality of health conditions} such as "ADHD," "Schizophrenia," "Autism," "Obesity," and "Type 2 diabetes." Their inclusion in the maximum k-core highlights their critical role within the research domain. These conditions are deeply interconnected, suggesting that they are essential to understanding ADHD, either as comorbidities, risk factors, or consequences. The presence of "Major depression" and "Coronary artery disease" further emphasizes the complex and multifaceted nature of ADHD, indicating its intersection with both mental and physical health domains.

Another significant finding is the \textbf{importance of cognitive and behavioral concepts}, such as "Intelligence," "Years of schooling," and "Subjective well-being," which are also part of the highest k-core. This underscores their significance in ADHD research, indicating that these factors are not peripheral but central to understanding and managing ADHD. The persistence of these nodes in the maximum k-core suggests that ADHD is intricately linked with broader cognitive and educational outcomes, which are crucial for both academic research and practical interventions.

Additionally, the \textbf{research focus on measurement and documentation} is evident, with nodes related to empirical studies and outcomes, such as "FMRI studies of cognitive control," "Body mass index," and "Smoking initiation," being present in the maximum k-core. This emphasizes the importance of empirical data and research documentation in ADHD studies. The inclusion of these nodes suggests that the methodologies and data sources used in ADHD research are fundamental to the domain, providing the empirical foundation upon which other concepts and conditions are built.

Finally, the \textbf{thematic grouping and network robustness} observed in the k-core analysis highlight the solid structure of the ADHD KG. The clustering of related concepts, such as health conditions and cognitive factors, within the maximum k-core, indicates that the graph is well-structured around these central themes, which are likely to be resilient to changes or perturbations. The robustness of the k-core structure, with a high \( k \) value, reflects the strength of the interconnections among these core nodes, indicating that the KG has a solid and reliable foundation based on Dr. Faraone's comprehensive research.

Overall, the k-core analysis of the ADHD KG reveals the most central and interconnected concepts within the domain, emphasizing the importance of specific health conditions, cognitive factors, and research methodologies. The high k-core value highlights the robustness of the graph's structure, indicating that these core concepts are deeply embedded in the research landscape of ADHD. This analysis not only enhances our understanding of the KG but also provides valuable guidance for future research and interventions in the field of ADHD.


\vspace{-3mm}
\section{Discussion}
\vspace{-2mm}

From the KG, we can see that, even with the relatively few documents provided for this preliminary work, the graph is huge. The graph will likely get much larger when provided with a more comprehensive set of documents. One way to reduce the complexity of the document will be to eliminate the clusters corresponding to persons and locations. These occur in the KG because some of the documents provided include names of authors and their institutions and geographical locations. Excluding this information would not reduce the accuracy of the graph but would reduce computational complexity. Likewise, dates appear as a cluster in the graph because article citations include dates but these are also not relevant to the goals of this project. These considerations also apply to the community detection analyses which show many nodes corresponding to dates, persons and locations. Deleting these nodes would dramatically reduce the number of edges as well.

The k-core analysis is instructive. It eliminates the issue with dates, persons and locations and highlights concepts that are mostly relevant to ADHD. Yet, some of the k-cores are either not relevant or only peripherally related to ADHD. “Number of Children” emerges as a k-core but is not useful for understanding ADHD or creating a chatbot. The k-cores also do not distinguish between concepts that have a strong connection to ADHD in the research literature (e.g., depression) from those that have a weak (schizophrenia) or questionable (lung cancer) relationship. It is also odd that two of the k-cores refer to methods (fMRI studies, genetic correlation). Further work is also needed to understand why the k-cores do not reference treatment approaches.

\vspace{-3mm}
\section{Conclusion}
\vspace{-2mm}
In this study, we applied network analysis to Attention-Deficit/Hyperactivity Disorder (ADHD) by constructing a multimodal KG that integrates data from diverse sources and expert knowledge. This structured representation maps intricate relationships between various factors associated with ADHD and offers deeper insights into its etiology and manifestation. Dr. Faraone’s extensive research was instrumental in informing our approach, providing validated measures that enhanced the relevance and robustness of our study. By building upon his contributions, we grounded our analysis in the most authoritative understanding of ADHD, further strengthening the credibility of our findings.

Our methodology leverages advanced techniques like contextual proximity calculations and node embeddings using Large Language Models (LLMs) to extract and structure ADHD-related knowledge. This approach reduced redundancy, captured subtle semantic connections, and employed clustering techniques to merge semantically similar concepts, resulting in a streamlined and nuanced KG. Despite using a relatively small dataset, the graph uncovered an extensive network of ADHD-related concepts, revealing its broad connections across various topics and disciplines. The graph’s potential for expansion further underscores its value as a research tool.

We also identified areas for improvement, such as eliminating irrelevant nodes to enhance interpretability and reduce computational complexity. This refinement process highlighted challenges in data extraction and preprocessing, suggesting areas for methodological improvements to optimize the graph further. The KG maps ADHD’s interdisciplinary connections, revealing intersections with conditions like depression, schizophrenia, and unexpected areas like lung cancer. Additionally, the absence of treatment-related K-cores points to gaps in the dataset, suggesting that therapeutic strategies may be underrepresented, requiring more refined techniques to distinguish core ADHD concepts.

By refining the graph, it can serve as a foundation for developing tools like targeted chatbots to provide valuable information to patients, clinicians, and educators. This comprehensive resource also facilitates interdisciplinary collaboration, supporting both future research and practical applications, such as enhancing clinical decision-making and AI-driven tools like Retrieval Augmented Generation (RAG) systems for ADHD-Expert LLMs.

This preliminary work uncovers the richness and complexity of ADHD knowledge and lays the groundwork for further research. Addressing current limitations and refining the graph will enhance its accuracy and usefulness, contributing to improved outcomes in the diagnosis, treatment, and management of ADHD.

%
\vspace{-3mm}
\bibliographystyle{spmpsci} 
\bibliography{paper} 



\end{document}